\documentclass[12pt]{article}

\usepackage{graphicx}
\usepackage{latexsym,array}
\usepackage{amsmath,amssymb}
\usepackage{bm}

\begin{document}

\begin{titlepage}

\begin{flushright}
\begin{tabular}{l}
KEK-TH-1347
\\
January 9, 2010
\\
Revised: February 16, 2010
\end{tabular}
\end{flushright}

\vspace{1\baselineskip}

\vspace*{5mm}

\begin{center}
{\Large \textbf{Lepton electric dipole moments 
in supersymmetric type II seesaw model} }

\vspace{0.5cm} 

{\large
Toru Goto\,$^{a}$, Takayuki Kubo\,$^{a,b}$ and Yasuhiro Okada\,$^{a,b}$
}\\[0.4cm]

$^{a}$\textit{
    KEK Theory Center, Institute of Particle and Nuclear Studies, KEK,
    Tsukuba, Ibaraki 305-0801, Japan
}\\[0.3cm]

$^{b}$\textit{
    The Graduate University for Advanced Studies (Sokendai), \\
    Tsukuba, Ibaraki 305-0801, Japan
}\\[0.3cm]

\vspace*{0.3cm} 

\begin{abstract}
We study the lepton electric dipole moments in the framework of the
supersymmetric type II seesaw model where the exchange of heavy
$SU(2)_{\mathrm{W}}$ triplets generates small neutrino masses.
We show that the CP violating phase of the bilinear soft supersymmetry
breaking term associated with the $SU(2)_{\mathrm{W}}$ triplets
contributes to lepton electric dipole moments mainly through threshold
corrections to the gaugino masses at the seesaw scale.
As a consequence, the ratio of the electric dipole moments of the muon
and the electron is the same as the ratio of their masses in a wide
region of parameter space.
\end{abstract}

\end{center}
\end{titlepage}

\newpage 

The electric dipole moments (EDMs) of leptons, nucleons and atoms are
important probe for physics beyond the Standard Model.
Until now, however, no EDMs has been observed experimentally and only
upper limits of them have been obtained
\cite{Regan:2002ta,EDMs-experimental-limits}.
It is known that the upper limits on EDMs strongly constrain CP
violating parameters of the new physics sector.
In particular,
supersymmetric (SUSY) extensions of the Standard Model are the most
promising candidates for the physics beyond the Standard Model.
The minimal version of them are usually referred to as the Minimal
Supersymmetric Standard Model (MSSM).
In the framework of the MSSM, new sources of CP violation are introduced
in terms of soft SUSY-breaking parameters and are severely constrained
by the upper limits on EDMs \cite{EDMs-in-SUSY}.

Recent neutrino observations tell us the neutrinos have nonzero but tiny
masses.
The seesaw mechanism \cite{seesaw-mechanism} is an attractive idea to
explain their smallness,
in which their masses are generated through the exchange of heavy
fields.
There are three types of seesaw models depending on the nature of the
heavy fields.
In the type I model, three generations of gauge singlet fermions
(right-handed neutrinos) with large Majorana masses are introduced.
The type II model \cite{seesaw-typeII} includes heavy
$SU(2)_{\mathrm{W}}$ triplet scalar field(s).
Heavy $SU(2)_{\mathrm{W}}$ triplet fermions are introduced in the type
III model \cite{seesaw-typeIII}.

In the framework of SUSY seesaw models, newly introduced superpotential
and soft SUSY breaking terms associated with the heavy fields may
contain CP violating phases which contribute to the EDMs of leptons and
quarks.
Thus the current bounds and future measurements of the EDMs can provide
us with information of physics at the seesaw scale. 
The lepton EDMs in the framework of the SUSY type I seesaw models 
have been studied by many
authors~\cite{Romanino:2001zf,Farzan:2003gn,Farzan:2004qu}.
In particular, Farzan~\cite{Farzan:2003gn} studied effects of the
bilinear soft SUSY breaking terms ($B$ terms) of heavy right-handed
sneutrinos.
It was shown that CP violating imaginary part of the $B$ term
contributes to trilinear coupling ($A$ term) of slepton and Higgs fields
through the threshold correction at the seesaw scale, and eventually
generates lepton EDMs.
A similar effect in the SUSY type II seesaw model was studied by Chun,
Masiero, Rossi and Vempati \cite{Chun:2005qw}, where the $B$ term
of heavy $SU(2)_{\mathrm{W}}$ triplet fields contributes to the slepton
$A$ term in the same way as in the type I case.
However, the $B$ term affects not only the slepton $A$ term but also 
the $SU(2)_{\mathrm{W}}\times U(1)_{\mathrm{Y}}$ gaugino
masses in the context of the type II seesaw model \cite{Joaquim:2006uz}.

In this Letter, we study the effects of the triplet $B$ term on lepton
EDMs in the SUSY type II seesaw model taking account of the
contribution from the gaugino masses as well as that from the slepton
$A$ term.
We show that the former is more important than the latter.
As a result, the ratio of muon EDM to electron EDM is given by
$d_{\mu}/d_e \simeq m_{\mu}/m_e$ in a wide region of parameter space.
In Ref.~\cite{Chun:2005qw}, where only the effect from the slepton $A$
term is considered, it is argued that the lepton EDM ratios are
determined by the neutrino parameters.
Our result is different from that of Ref.~\cite{Chun:2005qw} because the
contribution from the gaugino masses turns out to be generally larger
than that from the slepton $A$ term.

First, we briefly review the SUSY type II seesaw
model.
We follow the conventions of the SUSY Les Houches Accord \cite{SLHA}
for the MSSM sector.
The superpotential of the model is given by
\begin{eqnarray}
W &=& \epsilon_{ab} 
      \biggl(   Y_E^{ij} H_1^a L_i^b \bar{E}_j 
              + Y_D^{ij} H_1^a Q_i^b \bar{D}_j 
              - Y_U^{ij} H_2^a Q_i^b \bar{U}_j  
              - \mu H_1^a H_2^b
\nonumber \\  
  &&          + \frac{1}{\sqrt{2}} Y_T^{ij} L_i^a T_1^{bc} L_j^c 
              + \frac{1}{\sqrt{2}} \lambda_1 H_1^a T_1^{bc} H_1^c 
              + \frac{1}{\sqrt{2}} \lambda_2 H_2^a T_2^{bc} H_2^c
      \biggr) \nonumber \\
  &&  +M_T \mathrm{tr}(T_1 T_2), 
\label{eq:superpotential}
\end{eqnarray}
where $Q_i^a$, $L_i^a$, $\bar{E}_i$, $\bar{D}_i$, $\bar{U}_i$, $H_1^a$
and $H_2^a$ denote chiral supermultiplets in the MSSM with the suffixes
$a,b,c=1,2$ and $i,j=1,2,3$ being $SU(2)_{\mathrm{W}}$ and generation
indices, respectively.
$T_1$ and $T_2$ are $SU(2)_{\mathrm{W}}$ triplets with hypercharge
$1$ and $-1$, respectively.
By rephasing and rotating the fields, we can take the basis that $Y_E$
is real and diagonal, $\lambda_{2}$ and $M_T$ are real, $\lambda_{1}$ is
complex and $Y_T$ is a complex symmetric matrix.
The relevant soft SUSY breaking terms are given by
\begin{eqnarray}
\mathcal{L}^{\mathrm{soft}} 
&=& -\epsilon_{ab} 
     \biggl( 
             A_E^{ij} H_1^a \tilde{L}_i^b \tilde{e}_{Rj}^* 
           + A_D^{ij} H_1^a \tilde{Q}_i^b \tilde{d}_{Rj}^* 
           - A_U^{ij} H_2^a \tilde{Q}_i^b \tilde{U}_{Rj}^*
           - B_H \mu H_1^a H_2^b
 \nonumber \\  
 &&        + \frac{1}{\sqrt{2}} A_T^{ij} \tilde{L}_i^a T_1^{bc} \tilde{L}_j^c
           + \frac{1}{\sqrt{2}} A_1 H_1^a T_1^{bc} H_1^c 
           + \frac{1}{\sqrt{2}} A_2 H_2^a T_2^{bc} H_2^c   
           + \mathrm{h.c.}
           \biggr)
\nonumber \\  
 && -\Bigl( M_T B_T \mathrm{tr}[T_1 T_2] + \mathrm{h.c.} \Bigr) 
    +
    \frac{1}{2}
    \Bigl(
    M_1\tilde{b}\tilde{b} +M_2\tilde{w}\tilde{w} + \mathrm{h.c.}
\Bigr)
\nonumber \\  
 &&    - (m_{\tilde{L}}^2)_{ij} \tilde{L}^*_{ia} \tilde{L}_j^a
    + \cdots .
\label{softSUSYbreaking}
\end{eqnarray}
Here, $H_{1,2}^a$ and $T_{1,2}^{bc}$ denote scalar components of the
chiral multiplets which are given by the same notations in
(\ref{eq:superpotential}).
$\tilde{Q}_i^a$, $\tilde{L}_i^a$, $\tilde{e}_{Ri}^*$, $\tilde{d}_{Ri}^*$
and $\tilde{u}_{Ri}^*$ are scalar components of $Q_i^a$, $L_i^a$,
$\bar{E}_i$, $\bar{D}_i$ and $\bar{U}_i$, respectively.
$\tilde{b}$ and $\tilde{w}$ are $U(1)_{\mathrm{Y}}$ and
$SU(2)_{\mathrm{W}}$ gaugino fields, respectively.
To avoid large flavor changing neutral current effects, we assume that
the soft SUSY breaking mass terms are universal at a high energy scale
$M_G=2\times 10^{16}\,\mathrm{GeV}$ and that the $A$ terms are
proportional to the corresponding Yukawa couplings
($A_E^{ij}=a_0 Y_E^{ij}$ \textit{etc.}) at $M_G$.
In the following, we denote the universal scalar mass by $m_0$ and 
the constant proportionality by $a_0$.
We also assume that gaugino masses are universal at $M_G$ and are given
by $m_{1/2}$.\footnote{
Strictly speaking, $M_G$ is not the ``GUT scale'' since the existences
of the $T_1$ and $T_2$ spoil the gauge coupling unification.
We take these boundary conditions for technical simplicity.
A model with grand unification constructed by embedding $T_{1,2}$ into
$SU(5)$ multiplets is considered in
Refs.~\cite{Rossi:2002zb,Joaquim:2006uz}.
}

Under these boundary conditions, there remain three CP violating phases
that contribute to the EDMs: phases of $\mu$, $a_0$ and $B_T$.
The phases of $m_{1/2}$ and $B_H \mu$ are rotated away without loss of
generality.
Different CP violating sources contribute to lepton EDMs in different
ways. 
Effects of the phases of $\mu$ and $a_0$ have been studied in detail in
the literature \cite{Pospelov:2005pr}.
Here we study the effect of $B_T$ as a new source of CP violation and
assume $\mu$ and $a_0$ to be real parameters. 

The tiny neutrino masses are generated through the exchange of the
triplet fields, which is given by
\begin{equation}
  (m_{\nu})_{ij} =
  \frac{\lambda_2}{M_T} \Bigl( \frac{v_2}{\sqrt{2}} \Bigr)^2 (Y_T)_{ij}.
\label{Eq:m_nu}
\end{equation}
$v_2$ is the vacuum expectation value of $H_2$ field.
The matrix $m_{\nu}$ can be diagonalized by the Maki-Nakagawa-Sakata
(MNS) matrix \cite{Maki:1962mu}:
\begin{equation}
  \mathrm{diag} ( m_{\nu_1}, m_{\nu_2}, m_{\nu_3} ) 
  = m_{\nu}^{\mathrm{diag}}
  = U_{\mathrm{MNS}}^T m_{\nu} U_{\mathrm{MNS}},
\end{equation}
where the matrix $U_{\mathrm{MNS}}$ is defined as
\begin{equation}
  U_{\mathrm{MNS}} =
  V \, \mathrm{diag} (e^{-i\frac{\phi}{2}},\,e^{-i\frac{\phi'}{2}},\,1),
\label{U=VP}
\end{equation}
where $\phi$ and $\phi'$ are CP violating Majorana phases and
$V$ is given by
\begin{equation}
V = 
\begin{pmatrix}
  c_{12}c_{13}
& c_{13}s_{12}
& s_{13}e^{-i\delta}
\\
  -s_{12}c_{23}-c_{12}s_{23}s_{13}e^{i\delta}
&  c_{12}c_{23}-s_{12}s_{23}s_{13}e^{i\delta}
&  c_{13}s_{23}
\\
   s_{12}s_{23}-c_{12}c_{23}s_{13}e^{i\delta}
& -c_{12}s_{23}-c_{23}s_{12}s_{13}e^{i\delta}
&  c_{13}c_{23}
\end{pmatrix}.
\end{equation}
We have abbreviated $\sin \theta_{ij}$ and $\cos \theta_{ij}$
as $s_{ij}$ and $c_{ij}$, respectively.  
Hereafter we use the following parameters:
\begin{eqnarray}
&& \Delta m_{21}^2 = m_{\nu_2}^2 - m_{\nu_1}^2 = 8.0 \times 10^{-5} \, \mathrm{eV}^2
\,,
\nonumber\\
&& |\Delta m_{32}^2| = |m_{\nu_3}^2 - m_{\nu_2}^2| = 2.5 \times 10^{-3} \, \mathrm{eV}^2
\,,
\nonumber\\
&& \sin\theta_{12} = 0.56 \,, \hspace{0.5cm}
   \sin\theta_{23} = 0.71 \,, \hspace{0.5cm} 
   \sin\theta_{13} = 0.01 \,,
\label{neutrinoparameters}
\end{eqnarray}
and assume that $m_{\nu_1} \sim 0\,\mathrm{eV}$ which corresponds to the
normal hierarchy of the neutrino masses.
We take the values of $\Delta m_{21}^2$, $|\Delta m_{32}^2|$,
$\sin\theta_{12}$ and $\sin\theta_{23}$ from Ref.~\cite{Strumia:2006db}.
Note here that $Y_T^{\dagger} Y_T $ can be written as follows,
\begin{equation}
  (Y_T^{\dagger} Y_T )_{ij}
  \simeq
  \left( \frac{0.01}{\lambda_2}    \right)^{2}
  \left( \frac{1+\tan^2\beta}{\tan^2\beta} \right)^{2}
  \left( \frac{|M_T|}{10^{13} \, \mathrm{GeV}} \right)^{2} 
  \left(
    \frac{\sum_k m_{\nu k}^2 U_{\mathrm{MNS}}^{ik} U_{\mathrm{MNS}}^{jk*}
    }{10^{-3} \, \mathrm{eV}^2}
  \right),
\label{Eq:YY}
\end{equation}
where we have substituted $246 \,\mathrm{GeV}$ into the vacuum
expectation value $v=\sqrt{v_1^2+v_2^2}$ and the $\tan\beta$ is defined
by the ratio of the vacuum expectation values of the two Higgs fields:
$\tan\beta=v_2/v_1$.
In the type II seesaw model, the Yukawa coupling $Y_T$ is directly
related to the neutrino masses and the MNS matrix in contrast to the
type I seesaw model.

We calculate lepton EDMs in the SUSY type II seesaw model with use of
the following procedure.
We solve the renormalization group equations for the parameters in the
SUSY type II seesaw model from $M_G$ scale to $M_T$ with the input
parameters $m_0$, $a_0$ and $m_{1/2}$ at $M_G$.\footnote{%
The renormalization group equations that we have used agree with the
relevant part of those given in Ref.~\cite{Borzumati:2009hu}.
}
Next, at $M_T$ scale, we calculate one-loop threshold corrections for
the matching of the parameters in the SUSY type II seesaw model
and those in the MSSM as an effective theory in the low energy
scale.
Then the renormalization group equations for the MSSM parameters are
solved down to the electroweak scale to evaluate masses and mixing
matrices of the SUSY particles.
Finally we calculate the lepton EDMs with chargino-sneutrino and
neutralino-charged slepton one-loop diagrams.

\begin{figure}[tbph]
\centering
\includegraphics[width=5.5cm,clip]{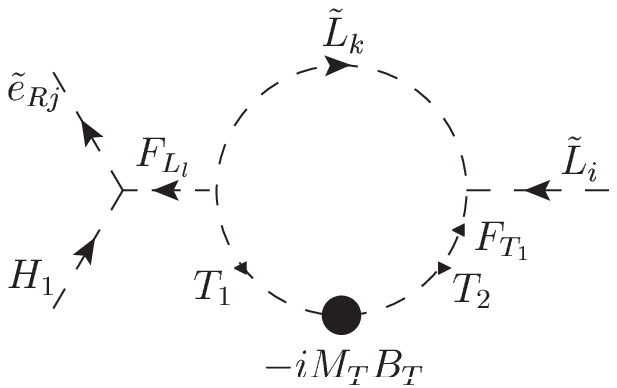}
\includegraphics[width=5.5cm,clip]{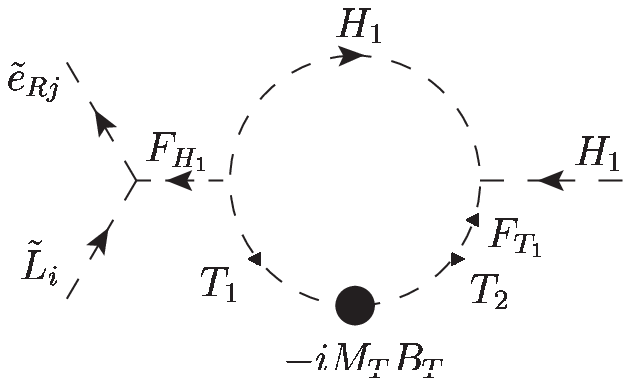}
\caption{%
One-loop Feynman diagrams that contribute to $\delta A_E$.
Fields $F_{T_1,H_1,L_i}$ represent auxiliary fields of supermultiplets.
}   
\label{fig:Aterm}
\end{figure}
\begin{figure}[tbph]
\centering
\includegraphics[width=5.5cm,clip]{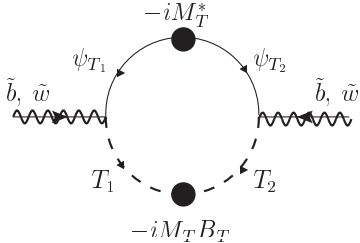}
\caption{%
One-loop Feynman diagram which gives threshold corrections to the
gaugino masses.
}   
\label{fig:gaugino}
\end{figure}

The $B_T$ term contributes to the threshold corrections only at the
$M_T$ scale where the triplet fields $T_{1,2}$ are integrated out.
There are two main contributions from $B_T$: threshold corrections
to the $A$ terms and those to the gaugino masses.
The threshold correction to the slepton $A$ term, denoted by $\delta A_E$,
is generated  through the diagrams shown in Fig.~\ref{fig:Aterm}. 
Keeping auxiliary fields as independent fields \cite{Farzan:2004qu}, 
we can easily calculate the correction $\delta A_E$:
\begin{equation}
\delta A_E
= \frac{3}{16\pi^2} B_T ( Y_T Y_T^{\dagger} + |\lambda_1|^2 ) Y_E.
\label{deltaAE_BT}
\end{equation}
Since $T_1$ and $T_2$ fields have
$SU(2)_{\mathrm{W}}\times U(1)_{\mathrm{Y}}$ gauge charges, threshold
corrections to the electroweak gaugino masses $M_{1,2}$ are generated.
One-loop correction terms proportional to $B_T$ are induced by the
diagram shown in Fig.~\ref{fig:gaugino}.
We obtain the threshold corrections $\delta M_1$ and $\delta M_2$ as
\begin{eqnarray}
\delta M_1 &=& -\frac{6}{16\pi^2} g^{\prime 2} B_T,
\label{deltaM1}
\\
\delta M_2 &=& -\frac{4}{16\pi^2} g^{2} B_T.
\end{eqnarray}
CP violating imaginary part of $B_T$ contributes to the lepton EDMs
through these threshold corrections.

\begin{figure}[tbph]
\centering
\begin{tabular}{cc}
\includegraphics[width=5.5cm,clip]{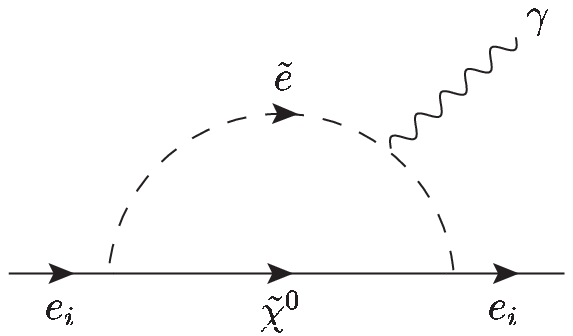} &
\includegraphics[width=5.5cm,clip]{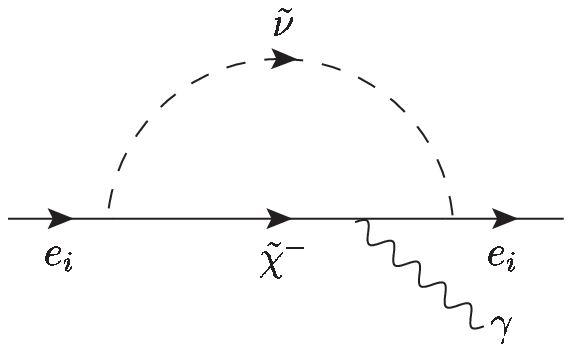} \\
(a) & (b)
\end{tabular}
\caption{%
One-loop diagrams contributing to lepton EDMs.
}
\label{fig:EDMmass*1} 
\end{figure}
Let us examine the effects
from $\delta A_E$ and $\delta M_{1,2}$ on the lepton EDMs.
Lepton EDMs are induced by the one-loop diagrams shown in
Fig.~\ref{fig:EDMmass*1}.
Contributions of $B_T$ through the neutralino-charged slepton diagram
(Fig.~\ref{fig:EDMmass*1}(a)) are schematically given by
\begin{eqnarray}
d_i^{\mathrm{Im}A_E} &\sim&
  \frac{e g^{\prime 2}}{(16\pi^2)^2}
  \frac{m_{e_i}}{m_{\mathrm{SUSY}}^4} 
  \mathrm{Re}M_1
  \bigl[ (Y_T Y_T^{\dagger})_{ii} + |\lambda_1|^2  \bigr] 
  \, \mathrm{Im}B_T,
\label{Eq:diImAE}
\\
d_i^{\mathrm{Im}M_1} &\sim&
  \frac{e g^{\prime 2}}{(16\pi^2)^2} \frac{m_{e_i}}{m_{\mathrm{SUSY}}^4} 
  (\mu g^{\prime 2} \tan\beta ) 
  \, \mathrm{Im}B_T,
\label{Eq:diImM1}
\end{eqnarray}
where $d_i^{\mathrm{Im}A_E}$ and $d_i^{\mathrm{Im}M_1}$ are
contributions of $\delta A_E$ and $\delta M_1$, respectively.
$m_{\mathrm{SUSY}}$ means a typical scale of SUSY particle masses in the
loop.
Contribution of $B_T$ in the chargino-sneutrino diagram
(Fig.~\ref{fig:EDMmass*1}(b)) through $\delta M_2$ is given by
\begin{eqnarray}
d_i^{\mathrm{Im}M_2} &\sim&
  \frac{e g^{2}}{(16\pi^2)^2} \frac{m_{e_i}}{m_{\mathrm{SUSY}}^4} 
  (\mu g^{2} \tan\beta ) 
  \, \mathrm{Im}B_T.
\label{Eq:diImM2}
\end{eqnarray}
We can see that the flavor dependence of the lepton EDMs induced by
(\ref{Eq:diImM1}), (\ref{Eq:diImM2}) and the term proportional to
$|\lambda_1|^2$ in (\ref{Eq:diImAE}) comes only from the lepton mass in
the overall factor.
On the other hand, the term proportional to $(Y_T Y_T^\dagger)_{ii}$ in
(\ref{Eq:diImAE}) has extra lepton flavor dependence determined
by the neutrino masses and mixings.
Therefore, if the contribution from $(Y_T Y_T^\dagger)_{ii}$ dominates,
the ratios of the lepton EDMs differ significantly from the
corresponding lepton mass ratios.
Otherwise, the lepton EDM ratios are approximately equal to the lepton
mass ratios.
In Ref.~\cite{Chun:2005qw}, it is argued that the ratios of EDMs are
given by 
\begin{equation}
\frac{d_i}{d_j} \simeq \frac{m_{e_i}}{m_{e_j}}
\frac{(Y_T Y_T^{\dagger})_{ii}}{(Y_T Y_T^{\dagger})_{jj}},
\label{ratioCMRV}
\end{equation}
taking the contribution from Eq.~(\ref{Eq:diImAE}) into account with the
assumption $|\lambda_1|^2 \ll Y_T Y_T^{\dagger}$.
If this relation is valid, we obtain $d_{\mu}/d_e \sim 10^4$ and
$d_{\tau}/d_{\mu} \sim 17$ substituting the neutrino parameters shown in
Eq.~(\ref{neutrinoparameters}) for the normal hierarchy of neutrino
masses.
However, since the contributions from Eq.~(\ref{Eq:diImM1}) and
Eq.~(\ref{Eq:diImM2}) are missing in Ref.~\cite{Chun:2005qw},
we calculate lepton EDMs including all the contributions in the
following.

In this model, it is known that the branching ratios of the lepton
flavor violating (LFV) processes such as $l_i \to l_j \gamma$ decays can
be large because of new source of lepton flavor mixing $Y_T$
\cite{Rossi:2002zb}.
Therefore, we calculate the branching ratios of
$l_i \to l_j \gamma$ as well as the lepton EDMs.
Since the $l_i \to l_j \gamma$ processes are induced by one-loop
diagrams of charginos (neutralinos) and sleptons, the branching ratios
$\mathrm{Br}(l_i \to l_j \gamma)$ are proportional to
$|(m_{\tilde{L}}^2)_{ij}|^2$.
In the SUSY type II seesaw model, the off-diagonal elements of
$m_{\tilde{L}}^2$ are mainly generated by the running between the $M_G$
and $M_T$ scales, which are roughly estimated as 
\begin{equation}
(m_{\tilde{L}}^2)_{ij}
\sim -\frac{m_0^2}{16\pi^2} (Y_T^{\dagger} Y_T)_{ij} \ln
\frac{M_G^2}{M_T^2}.
\label{offdiagonalm2L}
\end{equation}
In the numerical calculations, this effect is implicitly included in the
process of solving renormalization group equations.
We also take account of threshold corrections at $M_T$, which turn out
to be smaller than Eq.~(\ref{offdiagonalm2L}) by a
factor of $\ln (M_G / M_T)$.

\begin{figure}[tbph]
\centering
\begin{tabular}{cc}
\includegraphics[width=7cm,clip]{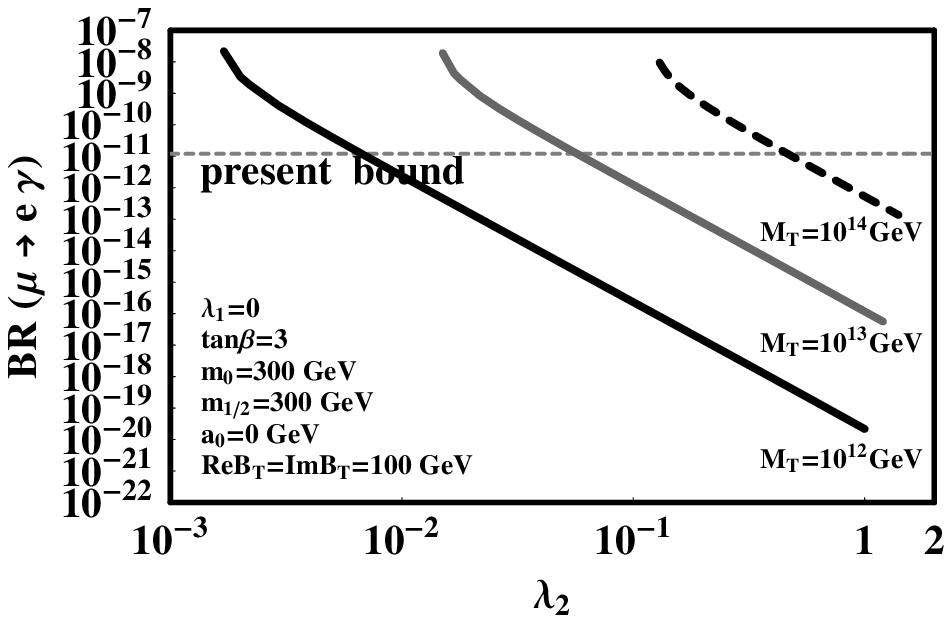} &
\includegraphics[width=7cm,clip]{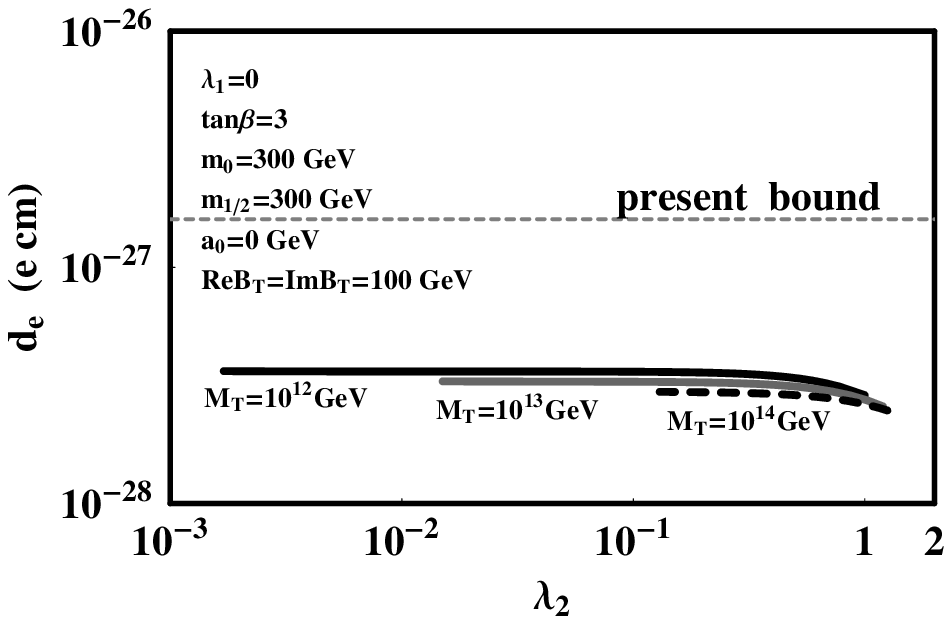} \\
(a) & (b) \\
\includegraphics[width=7cm,clip]{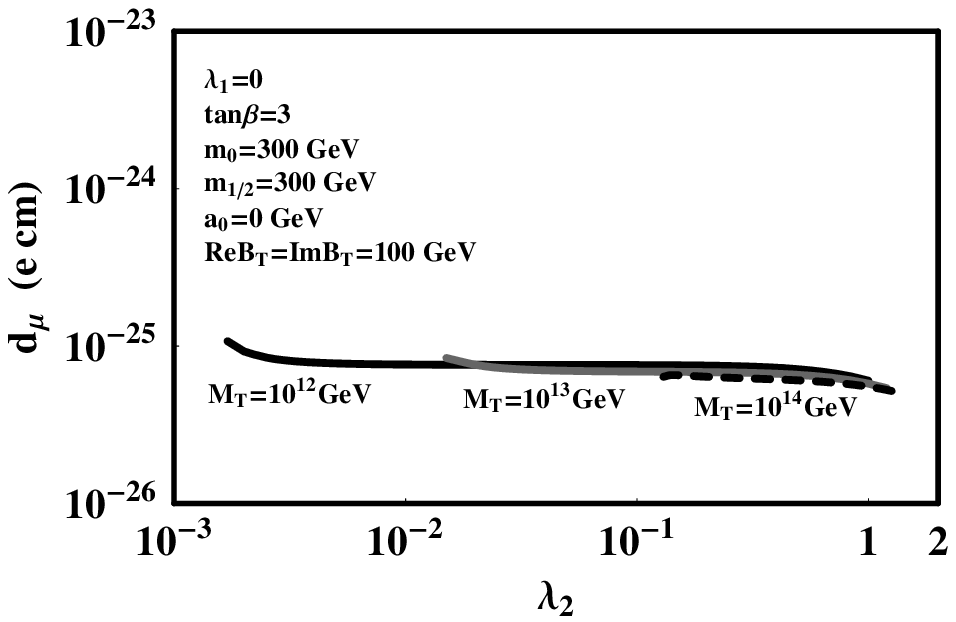} &
\includegraphics[width=7cm,clip]{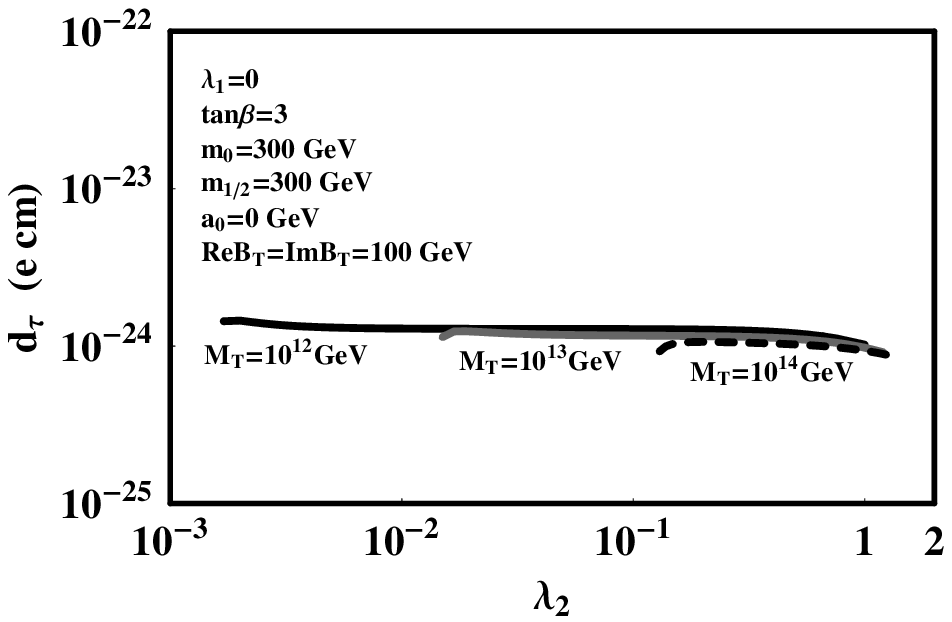} \\
(c) & (d)
\end{tabular}
\begin{tabular}{c}
 \includegraphics[width=7cm,clip]{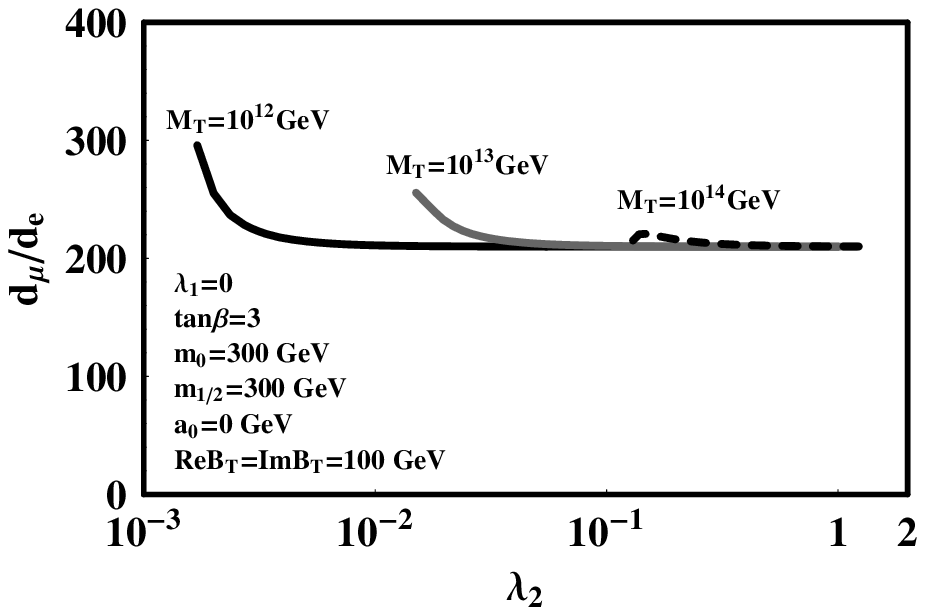}
\\
(e)
\end{tabular}
\caption{\footnotesize 
  (a) The branching ratio of $\mu\to e \gamma$, (b) the electron
  EDM, (c) the muon EDM, (d) the tau EDM and (e) the ratio of the muon
  EDM to the electron EDM as functions of $\lambda_2$
  for $\lambda_1=0$, $\tan\beta=3$, $m_0=m_{1/2}=300\,\mathrm{GeV}$,
  $a_0=0\,\mathrm{GeV}$ and
  $\mathrm{Re}B_T=\mathrm{Im}B_T=100\,\mathrm{GeV}$. 
  Black and gray solid lines and dashed lines are for $M_T=10^{12},
  10^{13}$ and $10^{14}\,\mathrm{GeV}$, respectively. 
  The input value of $\lambda_{1,2}$ and $B_T$ are given at the scale
  $M_T$ while those of $m_0$, $m_{1/2}$ and $a_0$ are given at the scale
  $M_G$.
}    
\label{fig:yokojiku_lambda2} 
\end{figure}

\begin{figure}[tbp]
\centering
\begin{tabular}{cc}
\includegraphics[width=6.7cm,clip]{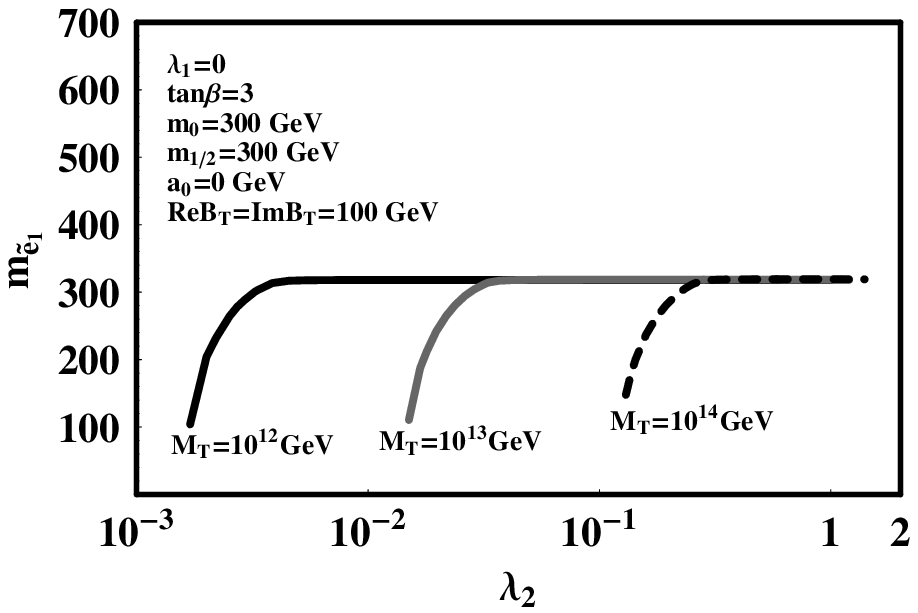} &
\includegraphics[width=6.7cm,clip]{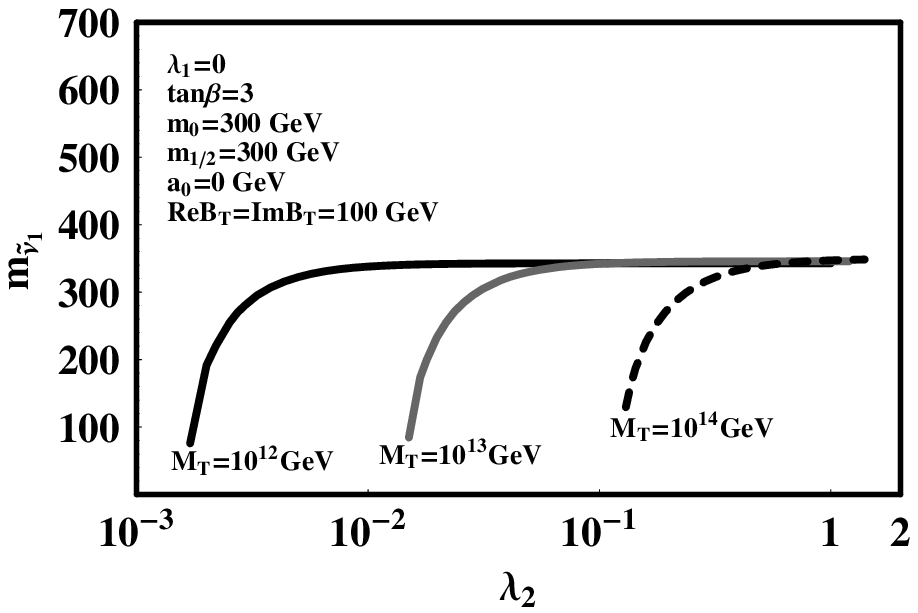}
\\
(a) & (b)
\end{tabular}
\caption{\footnotesize 
  Masses of the lightest charged slepton and the lightest sneutrino as
  functions of $\lambda_2$.
  Input parameters are the same as those in
  Fig.~\ref{fig:yokojiku_lambda2}.
}
\label{fig:yokojiku_lambda2_sleptonmass} 
\end{figure}

We show our numerical results of the branching ratio of
$\mu\to e \gamma$, the EDMs of electron ($d_e$), muon ($d_\mu$) and tau
($d_\tau$), and the ratio of $d_\mu$ and $d_e$ as functions of
$\lambda_2$ evaluated at $M_T$ scale for three cases of $M_T=10^{12}$,
$10^{13}$ and $10^{14}\,\mathrm{GeV}$ in
Fig.~\ref{fig:yokojiku_lambda2}.
We fix other input parameters as $\lambda_1=0$, $\tan\beta=3$,
$m_0=m_{1/2}=300\,\mathrm{GeV}$, $a_0=0\,\mathrm{GeV}$ and
$\mathrm{Re}B_T=\mathrm{Im}B_T=100\,\mathrm{GeV}$.
We take the Higgsino mass parameter $\mu$ as $\mu>0$.
In Figs.~\ref{fig:yokojiku_lambda2}(a) and
\ref{fig:yokojiku_lambda2}(b), current upper bounds of the branching
ratio of $\mu\to e\gamma$
$\mathrm{Br}(\mu \to e \gamma)<1.2\times10^{-11}$ \cite{Brooks:1999pu}
and the electron EDM $|d_e|<1.6\times 10^{-27}\,e\,\mathrm{cm}$
\cite{Regan:2002ta} are shown, respectively.
Since $Y_T$ is determined from $M_T$, $\lambda_2$ and the neutrino
parameters (\ref{neutrinoparameters}) by Eq.~(\ref{Eq:m_nu}),
a large (small) $\lambda_2$ corresponds to small (large) $Y_T$ for
a fixed $M_T$.
The lower limit of $\lambda_2$ in each plot is determined by the
conditions that $Y_T$ remains finite up to $M_G$ and the slepton masses
squared are positive at the electroweak scale.
The upper limit of $\lambda_2$ is set by the condition that $\lambda_2$
does not blow-up below $M_G$.
We can see that the ratio $d_{\mu}/d_e$ is around 200 except for the
lower end of $\lambda_2$ in each curve, but never becomes $10^4$ as
predicted in Eq.~(\ref{ratioCMRV}).
The reason why $d_{\mu}$ and $d_{\mu}/d_e$ grow at the smallest
values of $\lambda_2$ is that the mass of the lightest slepton
which couples to muon rather than electron rapidly decreases due to the
large $Y_T$ as shown in Fig.~\ref{fig:yokojiku_lambda2_sleptonmass}.  
This result implies that the contributions from $\delta M_{1,2}$ are
much larger than that from $\delta A_E$ in the whole parameter region%
\footnote{
We confirmed that the same results are obtained in the case we drop the
contribution from $\delta A_E$ by hand.
In the case we drop the contribution from $\delta M_{1,2}$ by hand, 
we can reproduce the results of the Ref.~\cite{Chun:2005qw}. 
}.
As seen in Fig.~\ref{fig:yokojiku_lambda2}(a),
$\mathrm{Br}(\mu \to e \gamma)$ exceeds the current experimental upper
limit in the region where $d_{\mu}/d_e$ deviates from the lepton mass
ratio $m_\mu/m_e$.
This is because $\mathrm{Br}(\mu \to e \gamma)$ is enhanced by
the large splitting among the slepton masses due to the large $Y_T$.
Consequently, after the experimental constraint on
$\mathrm{Br}(\mu \to e \gamma)$ is imposed, $d_{\mu}/d_e\simeq
m_\mu/m_e$ is satisfied in allowed parameter region.
As for $d_{\tau}$, we obtain
$d_{\tau}/d_{\mu}\simeq m_\tau/m_\mu\simeq 17$ in the whole parameter
region.
We also calculate the branching ratios of $\tau\to\mu\gamma$ and
$\tau\to e\gamma$.
We confirm that $\mathrm{Br}(l_i \to l_j \gamma)$ are controlled by the
neutrino parameters as discussed in
Refs.~\cite{Rossi:2002zb,Chun:2005qw,Joaquim:2006uz}, since the LFVs are
determined by $Y_T$ in this model.

\begin{figure}[tbph]
 \centering
 \begin{tabular}{cc}
 \includegraphics[width=7cm,clip]{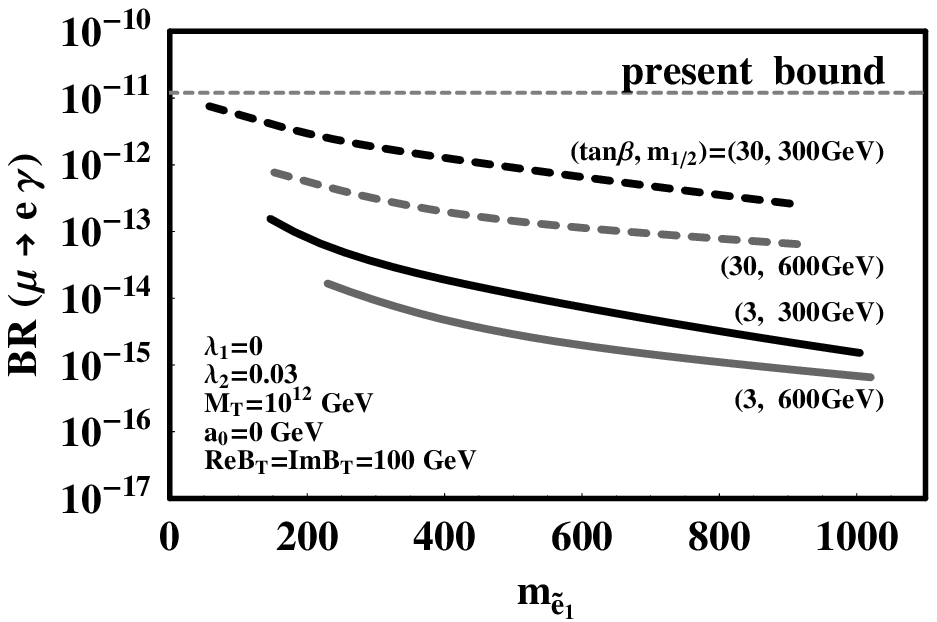} &
 \includegraphics[width=7cm,clip]{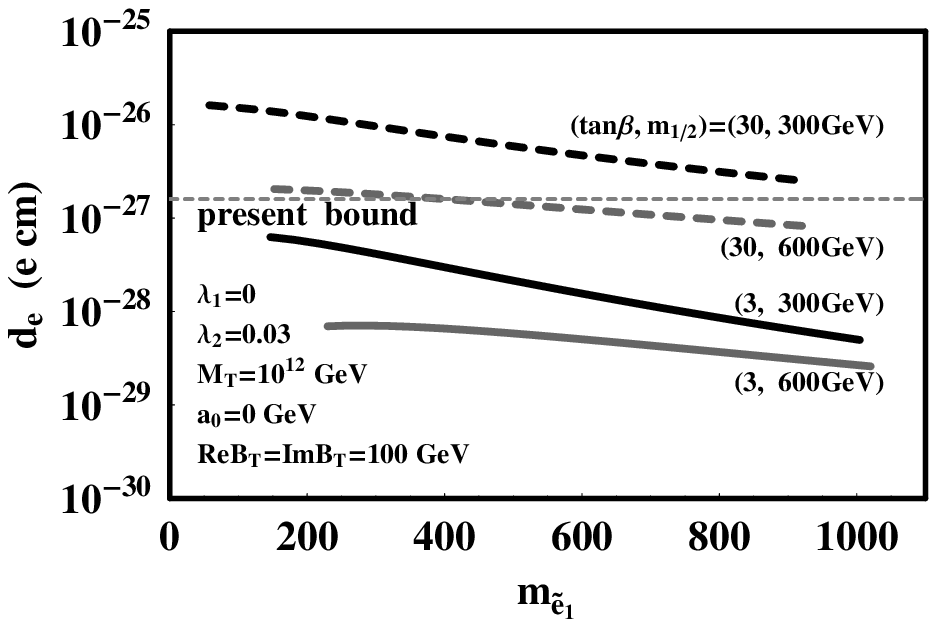} \\
(a) & (b) \\
 \includegraphics[width=7cm,clip]{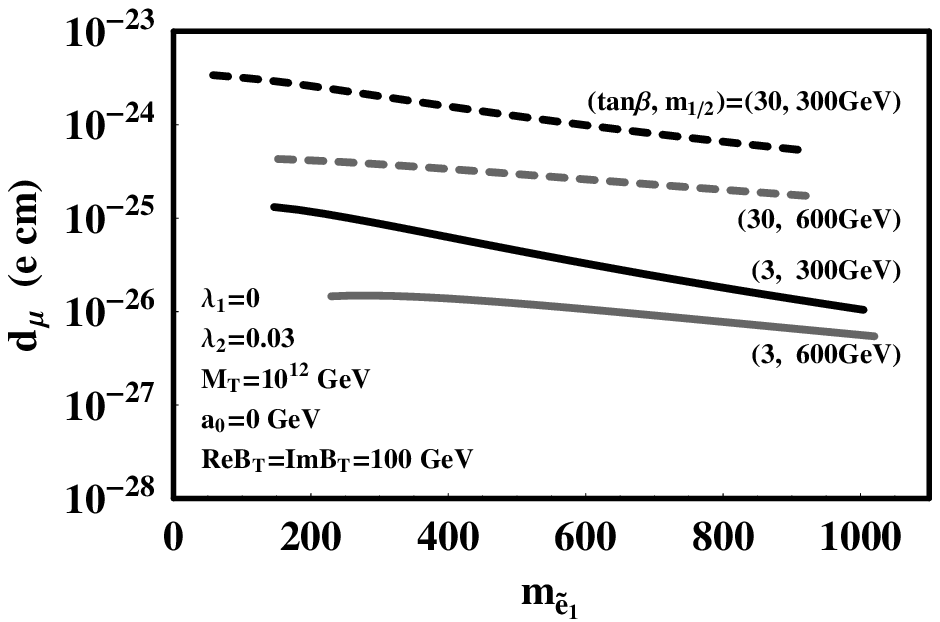} & 
 \includegraphics[width=7cm,clip]{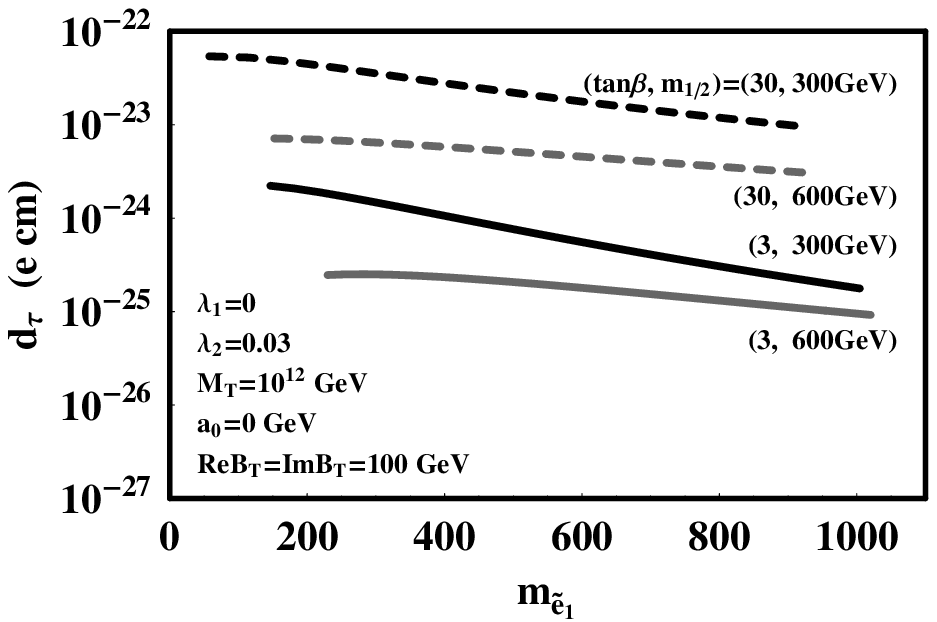} \\
(c) & (d)
\end{tabular}
\begin{tabular}{c}
 \includegraphics[width=7cm,clip]{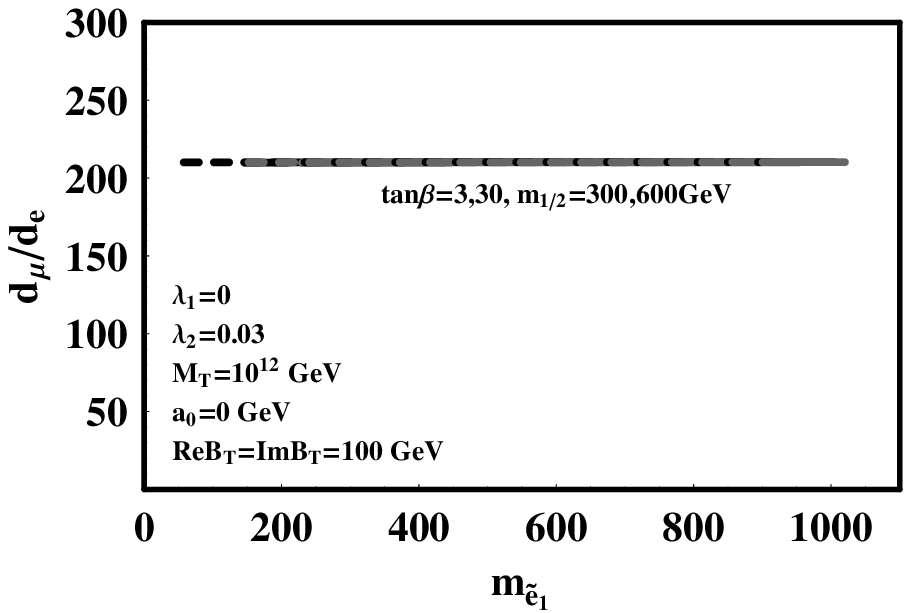}
 \\
(e)
 \end{tabular}
\caption{\footnotesize 
  (a) The branching ratio of $\mu\to e \gamma$,
  (b) the electron EDM,
  (c) the muon EDM,
  (d) the tau EDM and
  (e) the ratio of the muon EDM to the
  electron EDM as functions of the lightest charged slepton mass
  $m_{\tilde{e}_1}$ for $\lambda_1=0$, $\lambda_2=0.03$,
  $M_T=10^{12}\,\mathrm{GeV}$, $a_0=0\,\mathrm{GeV}$ and
  $\mathrm{Re}B_T=\mathrm{Im}B_T=100\,\mathrm{GeV}$.
  Black and gray solid lines are for
  $(\tan\beta, m_{1/2})=(3, 300\,\mathrm{GeV})$ and
  $(3, 600\,\mathrm{GeV})$, respectively, while black and gray dashed
  lines are for $(\tan\beta, m_{1/2})=(30, 300\,\mathrm{GeV})$ and
  $(30, 600\,\mathrm{GeV})$, respectively. 
}    
\label{fig:yokojiku_me1} 
\end{figure}

In Fig.~\ref{fig:yokojiku_me1}, we show the branching ratio of the
$\mu\to e \gamma$ decay, $d_e$, $d_\mu$, $d_\tau$ and $d_\mu/d_e$ as
functions of the lightest charged slepton mass $m_{\tilde{e}_1}$.
We vary $m_0$ within $100\,\mathrm{GeV}\leq m_0 \leq 1000\,\mathrm{GeV}$
and fix other parameters as $\lambda_1=0$, $\lambda_2=0.03$,
$M_T=10^{12}\,\mathrm{GeV}$, $a_0=0\,\mathrm{GeV}$,
$\mathrm{Re}B_T=\mathrm{Im}B_T=100\,\mathrm{GeV}$.
For $\tan\beta$ and $m_{1/2}$, we take the cases with $\tan\beta=3,\,30$
and $m_{1/2}=300,\,600\,\mathrm{GeV}$.
We see that the relation $d_\mu/d_e\simeq m_\mu/m_e$ holds in all cases.
\footnote{%
We have also calculated SUSY contribution to the muon anomalous magnetic
moment $a_\mu=(g_\mu-2)/2$.
Within the parameter space we have searched for, we obtain $0\lesssim
a_\mu(\mathrm{SUSY}) \lesssim 40\times 10^{-10}$ in the case of $\mu>0$.
This result is consistent with current experimental value
\cite{Davier:2009ag}.
}

In this Letter, we have studied leptonic EDMs in the SUSY type II seesaw
model including all contributions generated by one-loop threshold
corrections to SUSY breaking parameters at the seesaw scale through the
bilinear soft SUSY breaking term of the $SU(2)_{\mathrm{W}}$ triplet
fields.
We have shown that the ratios of the leptonic EDMs are given by those of
the lepton masses in a good approximation for most of parameter space.
We have presented numerical results for some specific cases, but this
conclusion holds unless fine tuning of parameters is made.
For instance, we have checked that the same conclusion is valid for the
case of $\lambda_1\neq 0$ or other types of neutrino mass hierarchy.
We have also relaxed the relation $M_1=M_2$ at the GUT scale and found
that the ratios of the EDMs do not change even if we varied $M_1/M_2$
within the range $1/10\leq M_1/M_2\leq 10$.
This result suggests that muon EDM is predicted to be 200 times larger
than the electron EDM in the SUSY type II seesaw model, which is
contrasted with the type I model where the relation is more complicated
because the $B$ term phase contributions to the EDMs depend on neutrino
Yukawa couplings \cite{Farzan:2003gn}.
Since the upper bound of the electron EDM is at the level of
$10^{-27}\,e\,\mathrm{cm}$, planned dedicated experiments
\cite{muEDMexp} of the muon EDM search at the level of
$10^{-24}-10^{-25}$ are very important.

The work of Y.\ O.\ is supported in part by the Grant-in-Aid for Science
Research, Ministry of Education, Culture, Sports, Science and
Technology, Japan, No.\ 16081211.
The work of T.\ G.\ and Y.\ O.\ is supported in part by the Grant-in-Aid
for Science Research, Japan Society for the Promotion of Science, No.\
20244037.

\end{document}